\documentclass{PoS}

\title{Rare tau decays at Belle}

\ShortTitle{Rare tau decays at Belle}

\author{\speaker{K.Hayasaka} for the Belle collaboration\\
        Kobayashi-Maskawa Institute, Nagoya University\\
        E-mail: \email{hayasaka@hepl.phys.nagoya-u.ac.jp}}

\abstract{We report results of a search for tau lepton decays 
strongly suppressed in the Standard Model based on the world-largest 
data sample accumulated with the Belle detector at the KEKB 
asymmetric-energy $e^+e^-$ collider. The decays include: 
lepton flavor and lepton number violating tau decays into 
a lepton ($e$ or $\mu$) and two charged mesons ($K$ or $\pi$)
as well as lepton and baryon number violating tau decays 
into a $\Lambda$ and a charged meson ($K$ or $\pi$). 
The sensitivity to the branching fractions is significantly
improved compared to our previous results and 
reaches ${\cal O}(10^{-8})$.}

\FullConference{The 2011 Europhysics Conference on High Energy Physics, EPS-HEP 2011,\\
		July 21-27, 2011\\
		Grenoble, Rh\^one-Alpes, France}

\begin{document}

\section{Introduction}
Generally, 
the tau lepton is expected to have a strong 
coupling to the new physics (NP)
since it has the heaviest mass among the leptons.
We search for many kinds of the signature for the NP
and lepton-flavor violation (LFV) is one of the
clear signatures 
because it is forbidden in the standard model (SM) and
its probability is very small even if the neutrino oscillations 
are taken into account.
Thus, a tau-lepton-flavor violating ($\tau$ LFV) decay
is promising and 
 many possible $\tau$ LFV modes are considered:
not only merely lepton-flavor-violating processes 
but also lepton-number- or baryon-number-violating processes.

Here, we report the recent results for $\tau\rightarrow\ell h h'$ 
and $\Lambda h$ using around a 1000
fb${}^{-1}$ data sample obtained by the Belle collaboration,
where $\ell=e, \mu$ and $h=\pi,K$.

\section{Method}

All searches for $\tau$ LFV decays follow  a similar procedure.
We search for $\tau^+\tau^-$ events
in which one $\tau$ (signal side) decays into an LFV mode under study,
while the other $\tau$ (tag side) decays
into one charged particle
and any number of additional photons and
neutrinos.
To search for exclusive LFV decay modes,
we select low multiplicity
events with zero net charge,
and separate an event into two hemispheres 
(signal and tag) using a thrust axis.
The backgrounds in such searches are dominated by
$q\bar{q}$ ($q=u,d,s,c$), generic $\tau^+\tau^-$, two-photon,
$\mu^+\mu^-$ and Bhabha events.
To obtain a good sensitivity,
we optimize the event selection
using particle identification and
kinematic information for each mode separately.
Because every $\tau$ LFV decay is a neutrinoless one,
the information from missing tracks is very powerful
to reject the background from generic $\tau^+\tau^-$
events.
After signal selection criteria are applied,
signal candidates are examined in the two-dimensional
space of the invariant mass, $M_{\rm {inv}}$, and the difference of
their energy from the beam energy in the CM system,
$\Delta E$. A signal event should have $M_{\rm {inv}}$
close to the $\tau$-lepton mass and $\Delta E$ close to 0.
We blind a region around the signal region
in the $M_{\rm{inv}}-\Delta E$ plane
so as not to bias our choice of selection criteria.
The expected number of background events in the blind region is first
evaluated,
and then the blind region is opened and candidate events are counted.
By comparing the expected and observed numbers of events,
we either observe  a $\tau$ LFV decay or set an upper limit (UL)
at 90\% confidence level (CL)
by applying counting approach.~\cite{counting}

\section{Results}

\subsection{\boldmath $\tau\rightarrow\ell h h'$ $(\ell=e, \mu$,
  $h=\pi,K)$}
Although $\tau^-\rightarrow\ell^-h^+ h^{\prime-}$
violates only the lepton flavor,
$\tau^-\rightarrow\ell^+h^- h^{\prime-}$
violates the lepton number as well.
Here, we search for 8 modes of  $\tau^-\rightarrow\ell^-h^+ h^{\prime-}$
and 6 modes of $\tau^-\rightarrow\ell^+h^- h^{\prime-}$.
The former ones are expected to be enhanced by the Higgs-mediated model
while the latter are motivated by the Majorana neutrino model.

We have updated the analysis for these modes with 
an 854 fb${}^{-1}$ data sample. Main backgrounds
come from $\tau\rightarrow\pi\pi\pi\nu$ for $\ell=\mu$
and $\tau\rightarrow\pi\pi^0\nu$ for $\ell=e$,
where the photon, that is a daughter of $\pi^0$,
converts to $e^+e^-$. To reduce them, 
$\tau\rightarrow\pi\pi\pi\nu$ veto, using $M_{\pi\pi\pi}$,
and $\gamma$-conversion veto are newly introduced
for $\mu\pi K$ modes and $\ell=e$ modes, respectively.
As a result, we have found 1 event for $\mu^+\pi^-\pi^-$
and $\mu^-\pi^+K^-$ modes while no events are observed
in the other modes. 
Since this is consistent with the expected number
of the backgrounds, we set the UL on the branching
fraction. 
The evaluated numbers to set the UL
and the number of observed events 
in the signal region 
are summarized in Table~\ref{tbl:eff2}.
The ULs for this mode are the most sensitive.
(preliminary)
\begin{figure}[h]
 \begin{center}
\includegraphics[width=0.9\textwidth]{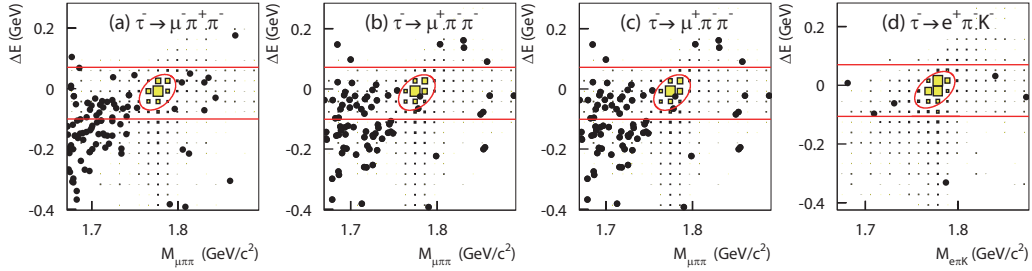}
\caption{\label{lhh}\small Resulting plots on the $M_{\rm inv}$ -- $\Delta E$
  plane
for $\tau^-\rightarrow\mu^-\pi^+\pi-$ (a),
  $\tau^-\rightarrow\mu^+\pi^-\pi^-$ (b),
  $\tau^-\rightarrow\mu^-\pi^+K^-$ (c) and 
  $\tau^-\rightarrow e^+ \pi^- K^-$ (d). Here, black dots show
the data and yellow boxes indicate the density for the signal events.}
 \end{center}
\end{figure}

\begin{table}[t]
\begin{center}
\caption{\label{tbl:eff2}\small
Summary of ULs for recently updated modes.
The signal efficiency~($\varepsilon$), 
the number of expected background events~($N_{\rm BG}$)
estimated from the  sideband data, 
the number of observed events 
in the signal region~($N_{\rm obs}$)
and 90\% CL UL on 
the branching  fraction~(${\cal{B}}_{90}$)
for each individual mode. }
\small
\begin{tabular}{c|cccc||c|cccc}\hline \hline
Mode & & & &  ${\cal{B}}_{90}$&
Mode & & & &  ${\cal{B}}_{90}$\\
($\tau^-\rightarrow$)&  $\varepsilon$~{(\%)} & 
$N_{\rm BG}$ &
 $N_{\rm obs}$ &
$(10^{-8})$ 
&
($\tau^-\rightarrow$)&  $\varepsilon$~{(\%)} & 
$N_{\rm BG}$ &
 $N_{\rm obs}$ &
 $(10^{-8})$ 
\\ \hline

$ \mu^-\pi^+\pi^- $ & 5.83 & $0.63\pm{0.23}$ 
 & 0 & 2.1 & 
$ e^-\pi^+\pi^- $ & 5.45 & $0.55\pm{0.23}$ 
 & 0 & 2.3\\
$ \mu^+\pi^-\pi^- $ & 6.55 & $0.33\pm{0.16}$ 
 & 1 &  3.9 &
$ e^+\pi^-\pi^- $ & 6.56 & $0.37\pm{0.18}$
 & 0 & 2.0\\

$ \mu^-K^+K^- $ & 2.85 & $0.51\pm{0.18}$ 
 & 0 & 4.4&
$ e^-K^+K^- $ & 4.29 & $0.17\pm{0.10}$ 
 & 0 & 3.4\\
$ \mu^+K^-K^- $ & 2.98 & $0.25\pm{0.13}$ 
 & 0 & 4.7&
$ e^+K^-K^- $ & 4.64 & $0.06\pm{0.06}$ 
 & 0 & 3.3\\

$ \mu^-\pi^+K^- $ & 2.72 & $0.72\pm{0.27}$
 & 1 &  8.6&
$  e^-\pi^+K^- $ & 3.97 & $0.18\pm{0.13}$
 & 0 & 3.7\\

$ \mu^-K^+\pi^- $ & 2.62 & $0.64\pm{0.23}$
 & 0 & 4.5&
$  e^-K^+\pi^- $ & 4.07 & $0.55\pm{0.31}$ 
 & 0 & 3.1\\

$ \mu^+K^-\pi^- $ & 2.55 & $0.56\pm{0.21}$
 & 0 & 4.8 &
$  e^+K^-\pi^- $ & 4.00 & $0.46\pm{0.21}$ 
 & 0 & 3.2 
 \\
\hline
$ {\Lambda}\pi^- $ 
& 4.39 & $0.31\pm{0.18}$ 
 & 0 & 3.0 &
$ {\Lambda}K^- $ 
& 3.16 & $0.42\pm{0.19}$ 
 & 0  & 4.2 \\
$ \bar{\Lambda}\pi^- $ 
& 4.80 & $0.21\pm{0.15}$ 
 & 0  & 2.8 &
$ \bar{\Lambda}K^- $ 
& 4.11 & $0.31\pm{0.14}$ 
 & 0  & 3.1 \\

\hline\hline
\end{tabular}
\end{center}
\end{table}

\subsection{\boldmath $\tau\rightarrow\Lambda h $ $(h=\pi,K)$}
This mode violates the lepton number $(L)$ as well as
the baryon number $(B)$:
$\tau^-\rightarrow \bar{\Lambda} h^- $ conserves 
$(B-L)$ while $\tau^-\rightarrow {\Lambda} h^- $ violates
$(B-L)$.
Some GUTs can make $(B-L)$ conserving decays while 
a more complicated model is necessary
 to induce $(B-L)$ violating decays.

We have searched for these decays with a 906 fb${}^{-1}$ data sample.
Main backgrounds come from $\tau\rightarrow\pi K_{S}^0 \nu$ and
$q\bar{q}$ ($q=u,d,s$) events having $\Lambda$ and $\pi$.
In the former background, $K_{S}^0$ is misidentified as $\Lambda$.
This can be rejected by a $K_S^0$ veto using $M_{\pi\pi}$.
On the other hand, the latter is likely to have a proton
as a charged track on the tag side
because of the baryon number conservation.
Therefore, we veto the proton for the tag-side track.
As a result, no events are observed for each mode;
no excess is found for the signal.
Thus, we set the UL on the branching
fraction. The obtained numbers are summarized
in Table~\ref{tbl:eff2}.
We obtain the most stringent ULs.
(preliminary)

\begin{figure}[h]
 \begin{center}
\includegraphics[width=0.9\textwidth]{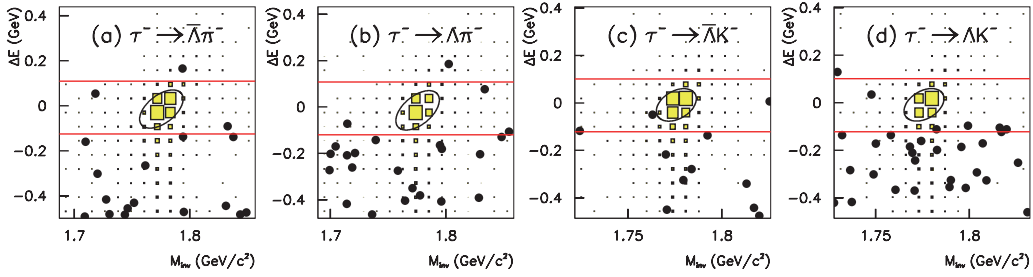}
\caption{\label{Lh}\small Resulting plots on the $M_{\rm inv}$ -- $\Delta E$
  plane
for $\tau^-\rightarrow\mu^-\pi^+\pi^-$ (a),
  $\tau^-\rightarrow\mu^+\pi^-\pi^-$ (b),
  $\tau^-\rightarrow\mu^-\pi^+K^-$ (c) and 
  $\tau^-\rightarrow e^+ \pi^- K^-$ (d). Here, black dots show
the data and yellow boxes indicate the density for the signal events.}
 \end{center}
\end{figure}

\section{Summary}

\begin{figure}[h]
\begin{center}
\includegraphics[width=12cm]{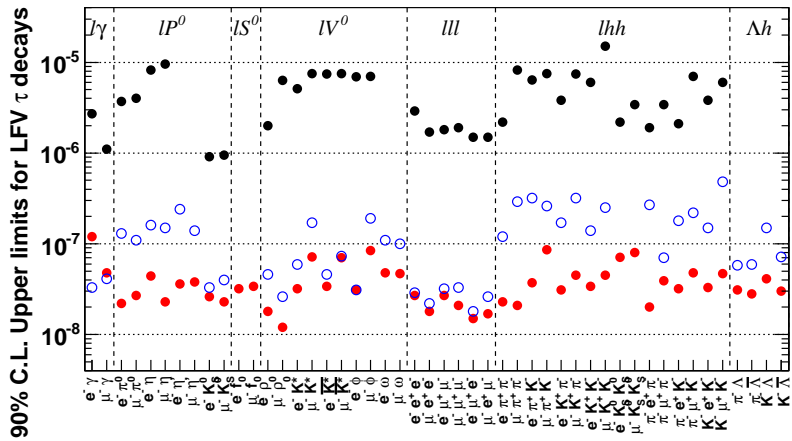}
\caption{\label{uls}\small
Current 90\% CL ULs for the 
branching fraction of $\tau$ LFV mode. Red, blue and black
circles show Belle, BaBar and CLEO results, respectively.}
\end{center}
\end{figure}

We have searched for
46 major modes of lepton flavor violating
$\tau$ decays with a 1000 fb${}^{-1}$ data sample,
reported the recently updated results 
for $\tau\rightarrow\ell h h'$ and $\Lambda h$ here
and obtained the most sensitive results for them so far.
The current status of the $\tau$ LFV searches in $B-$factory experiments
is summarized in Fig.~\ref{uls}.
No evidence for these decays is observed and
we set 90\% CL ULs
on the branching fractions at the $O(10^{-8})$ level.
The sensitivity for the LFV search is 100 times improved
in comparison with CLEO's one due to the effective background
rejection and increase of the data sample.

\end{document}